\documentclass{article}
\usepackage{graphicx} % Required for inserting images
\usepackage{amsmath}
\usepackage{amssymb}
\usepackage{amsthm}
\usepackage{physics}
\usepackage{authblk}
\usepackage{xcolor}
\usepackage{tikz}
\usetikzlibrary{positioning}
\usepackage{yquant}

\usepackage[style=numeric-comp, sorting=none]{biblatex}

\newtheorem{theorem}{Theorem}
\newtheorem{corollary}{Corollary}

\theoremstyle{definition}
\newtheorem{definition}{Definition}

\newcommand{\E}[2][]{\mathbb{E}_{#1}\left[#2\right]}
\allowdisplaybreaks[1]

\bibliography{references}

\title{Trainability of IQP Quantum Circuit Born Machines Under Gaussian Initialization}
\author[1]{Gennaro De Luca\thanks{gennaro.deluca@asu.edu}}
\author[1]{Vinayak Sharma}
\author[1]{Aviral Shrivastava}
\affil[1]{SCAI, Arizona State University}
\date{}

\begin{document}

\maketitle

\begin{abstract}
% Quantum Circuit Born Machines (QCBMs) instantiated with Instantaneous Quantum Polynomial (IQP) circuits can be optimized entirely classically by reformulating the Maximum Mean Discrepancy (MMD) loss via Pauli expectation values. 
% While this bypasses quantum hardware noise during training, optimization remains bottlenecked by barren plateaus and MMD kernel concentration. 
% In this work, we establish a rigorous theoretical framework for the trainability of IQP-QCBMs under Gaussian parameter initialization $\theta_j \sim \mathcal{N}(\mu_\theta, \sigma_\theta^2)$. 
% Using Stein's lemma we derive an analytical lower bound on MMD gradient variance as a function of $\exp(-2\sigma_\theta^2 |\Gamma(a)|)$, where $\Gamma(a) := \{j : g_j \cdot a = 1\}$ counts the active generators overlapping with observable \(a\).

% defines the reverse light cone of active gates. 
% This proves that gate connectivity rather than overall circuit depth governs gradient concentration. 
% To guarantee non-vanishing gradients, we establish exact initialization scaling rules: $\sigma_\theta^2 = \mathcal{O}(1 / \max_a |\Gamma(a)|)$, $\mu_\theta = \mathcal{O}(1 / \sqrt{\max_a |\Gamma(a)|})$, and MMD kernel bandwidth $\sigma^2 = \Omega(n)$. 
% Finally, we derive a Gaussian concentration bound formalizing gradient deviation probabilities across circuit topologies.

Quantum Circuit Born Machines (QCBMs) offer a natural approach to generative machine learning by leveraging the Born rule. Recent work demonstrates that QCBMs instantiated with Instantaneous Quantum Polynomial (IQP) circuits can be trained entirely classically via the Maximum Mean Discrepancy (MMD) loss. However, trainability remains constrained due to exponential concentration and barren plateaus. In this work, we leverage Stein's lemma and Lipschitz concentration bounds to establish a rigorous trainability framework for IQP-QCBMs initialized under Gaussian parameters $\theta_j \sim \mathcal{N}(\mu_\theta, \sigma_\theta^2)$. We derive an analytical lower bound on gradient variance governed by $\exp(-2\sigma_\theta^2 |\Gamma(a)|)$, showing that concentration depends on parameter-observable overlap counts $\Gamma(a) := \{j : g_j \cdot a = 1\}$ rather than geometric locality assumptions or circuit depth alone. We also show that scaling parameter initialization variance as $\sigma_\theta^2 = \mathcal{O}(1 / \max_a |\Gamma(a)|)$, mean as $\mu_\theta = \mathcal{O}(1 / \sqrt{\max_a |\Gamma(a)|})$, and MMD kernel bandwidth as $\sigma = \Omega(n)$ mitigates exponential concentration and prevents barren plateaus. Finally, we derive a probabilistic concentration bound on gradient deviations showing the conditions under which barren plateaus are likely to happen.

\end{abstract}

\section{Introduction}
Quantum Circuit Born Machines (QCBMs) offer a promising framework for generative quantum machine learning, leveraging the Born rule to define probability distributions through quantum state measurements \cite{liu2018differentiable, coyle2020born, gili2023quantum}. When instantiated with Instantaneous Quantum Polynomial (IQP) circuits \cite{bremner2016average}, QCBMs exhibit a powerful duality. Sampling from these circuits is provably hard for classical computers under standard complexity assumptions \cite{bremner2016average}, yet their Pauli expectation values can be calculated classically in polynomial time \cite{nest2009simulating}. By reformulating the Maximum Mean Discrepancy (MMD) loss \cite{liu2018differentiable} strictly in terms of these analytical expectation values, IQP-QCBMs can be optimized entirely classically \cite{recio2025train}. This "train classically, deploy quantumly" paradigm avoids shot noise from quantum measurement and physical device errors during optimization, allowing models to be trained classically prior to deploying quantum sampling at scale, making them exceptionally hardware-efficient and attractive for NISQ-era applications \cite{recio2025train}.

Despite these practical advantages, the optimization of quantum machine learning models remains severely constrained by trainability bottlenecks. Chief among these is the phenomenon of barren plateaus \cite{mcclean2018barren}, where the variance of cost function gradients vanishes exponentially with the increase in the number of qubits, leaving optimization algorithms stranded on flat cost landscapes. In loss formulations such as the MMD, this issue is intimately linked to kernel concentration \cite{thanasilp2024exponential}, where parameter updates become exponentially concentrated around their mean value, rendering the loss landscape uninformative and data-independent. Critically, these issues are inherent to the quantum feature space and therefore also plague classically trained IQP-QCBMs even in the absence of noise. 

To address the trainability challenges in the classically trained IQP-QCBMs, recent work \cite{lerch2026iqp} explored mitigation strategies like warm starts and data-dependent initializations, but their framework remains tied to uniform hypercube sampling. Beyond that, \cite{shen2026characterizing} analyzed trainability under uniform initialization ($\theta_j \sim U[0, 2\pi)$). However, uniform sampling of parameters across $[0, 2\pi)$ forces quantum states to globally over-explore Hilbert space, causing destructive phase interference that guarantees exponential gradient decay (barren plateaus) as system size grows. As such it is useful to expand this analysis to other parameter initialization schemes. 

As noted by \cite{shen2026characterizing} themselves,  that small-variance Gaussian initialization ($\theta_j \sim \mathcal{N}(\mu_\theta, \sigma_\theta^2)$) may hold significant promise in addressing the trainability of IQP-QCBMs. Analogous to variance-scaled initializations like Xavier \cite{glorot2010understanding} or He \cite{he2015delving} in classical deep learning, tuning $\sigma_\theta^2$ controls how quantum states explore Hilbert space—restricting trajectories to informative local sub-regions to prevent exponential concentration \cite{zhang2022escaping}. However, their Gaussian analysis relies on global worst-case assumptions, yielding loose bounds that miss how initialization variance $\sigma_\theta^2$ interacts with local gate coupling and phase cancellations.

This work extends the analysis of trainability of IQP-QCBMs under Gaussian parameter initialization and establishes a rigorous theoretical framework for IQP-QCBM trainability under Gaussian initializations. By leveraging Stein's lemma and Lipschitz concentration inequalities, our theoretical analysis establishes analytical lower bounds on MMD gradient variance and probabilistic concentration for IQP-QCBMs under Gaussian initializations. The main results of this paper are:
\begin{itemize}
    \item We derive an analytical closed-form expression for the MMD cost gradient, demonstrating that non-zero updates occur strictly for observables $Z_a$ that overlap with a parameter's generator mask ($g_k \cdot a = 1$). Consequently, sparse IQP generators inherently induce a sparse gradient structure.

    \item We establish an analytical lower bound on loss gradient variance showing gate cancellations restrict observable dependencies to a local reverse light cone, $\Gamma(a) := \{j : g_j \cdot a = 1\}$. Thus, gradient variance decays as $\exp(-2\sigma_\theta^2 \vert{}\Gamma(a)\vert{})$, showing that local gate connectivity -- rather than total circuit depth -- determines whether an IQP circuit concentrates.

    \item We prove that exponential gradient concentration is strictly mitigated when the parameter initialization variance scales as $\sigma_\theta^2 = \mathcal{O}(1/\max_a \vert{}\Gamma(a)\vert{})$, the parameter mean scales as $\mu_\theta = \mathcal{O}(1/\sqrt{\max_a \vert{}\Gamma(a)\vert{}})$, and the MMD kernel bandwidth scales as $\sigma = \Omega(n)$. Consequently, tuning initialization variance to the local reverse light cone size preserves trainability across arbitrary IQP circuit structures.

    \item We establish a Gaussian concentration bound for MMD cost gradients, showing that deviation probabilities are controlled by the product of the parameter variance and the Lipschitz constant of the gradient of the loss. This bound shows that a small variance and dense circuit can both diagnose the occurrence of a barren plateau.
\end{itemize}

\section{Preliminaries and Mathematical Setup}
Let $n$ denote the number of qubits, and $x, z \in \{0,1\}^n$ represent binary bitstrings. The bitwise inner product modulo 2 is $x \cdot z = \sum_{i=1}^n x_i z_i \pmod 2$.

\subsection{Quantum Circuit Born Machines (QCBMs)}
A Quantum Circuit Born Machine (QCBM) is an $n$-qubit generative model that approximates a target classical distribution $p(x)$ over bitstrings $x \in \{0,1\}^n$ using a parameterized quantum state $\vert\psi(\theta)\rangle$. According to the Born rule, measuring this state in the computational basis yields samples from the model distribution:

\begin{equation}
q_\theta(x) = |\langle x |\psi(\theta)\rangle|^2.
\end{equation}

Directly optimizing over full $2^n$-dimensional probability distributions is computationally intractable. Instead, we train the QCBM by matching expectation values of multi-qubit Pauli-$Z$ observables. These observables act as parity functions on subsets of the data bits, allowing the model to learn by matching these statistical features.

For a bitstring mask $a \in \{0,1\}^n$, the corresponding observable $Z_a = \bigotimes_{i=1}^n Z_i^{a_i}$ applies a Pauli-$Z$ operator to the $i$-th qubit if $a_i=1$, and the identity $I$ otherwise. This observable effectively extracts a parity feature from the data, and its expectation value over the target distribution $p$ is given by:

\begin{equation} \label{eq:data_exp}
\langle Z_a\rangle_p = \mathbb{E}_{x \sim p}\left[(-1)^{x \cdot a}\right].
\end{equation}

\begin{figure}[t!]
    \centering
    \definecolor{hadamardblue}{HTML}{DAE8FC}
    \definecolor{gyellow}{HTML}{FFF2CC}
    \definecolor{measurered}{HTML}{F8CECC}
    \begin{tikzpicture}
        % \yquantset{operator/separation=4mm}
        \begin{yquant}
            qubit {$\ket{0}$} j[3];
            [fill=hadamardblue] box {$H^{\otimes 3}$} (j);
            
            [fill=gyellow, label={above:$g_1$}] box {$e^{-i\frac{\theta_1}{2} \cdot (Z\otimes I \otimes I)}$} j[0];
            [fill=gyellow, label={below:$g_2$}] box {$e^{-i\frac{\theta_2}{2} \cdot (I\otimes Z \otimes Z)}$} (j[1],j[2]);
            [fill=gyellow, label={below:$g_3$}] box {$e^{-i\frac{\theta_3}{2} \cdot (I\otimes I \otimes Z)}$} (j[2]);
            [fill=hadamardblue] box {$H^{\otimes 3}$} (j);
            
            % Add sub figure here for measurement
            [fill=measurered] measure {$Z$} j[0];
            output {$I$} j[1];
            [fill=measurered] measure {$Z$} j[2];
        \end{yquant}
    \end{tikzpicture}
    \caption{IQP QCBM with the generator strings $g_j$~=~{[1,0,0], [0, 1, 1], [0, 0, 1]}, and measurement configuration a = [101]}
    \label{fig:iqp-qcbm}
\end{figure}

\subsection{Instantaneous Quantum Polynomial Circuits}
An $n$-qubit Instantaneous Quantum Polynomial (IQP) circuit generates the parameterized quantum state $\ket{\psi(\theta)}$ using diagonal $Z$-basis phase gates bounded by Hadamard layers:

\begin{equation}
\begin{aligned}
    \ket{\psi(\theta)} &= H^{\otimes n} V(\theta) H^{\otimes n} |0\rangle^{\otimes n}, \quad \text{where} \\
    V(\theta) &= \prod_{j=1}^m \exp\left(-i \frac{\theta_j}{2} Z_{g_j}\right).
\end{aligned}
\end{equation}

The circuit's architecture is defined by $m$ generator bitstrings $G = \{g_1, \dots, g_m\} \subset \{0,1\}^n$. Each bitstring $g_j$ acts as a mask determining which qubits are coupled by the trainable parameter $\theta_j$ (via the operator $Z_{g_j} = \bigotimes_{i=1}^n Z_i^{g_{j,i}}$). 

For example, in Figure \ref{fig:iqp-qcbm}, $G = \{[1,0,0], [0,1,1], [0,0,1]\}$ specifies the $Z$ generators $Z \otimes I \otimes I$, $I \otimes Z \otimes Z$, and $I \otimes I \otimes Z$. The overlap between any generator mask $g_j$ and a measurement mask $a$ (e.g., $Z \otimes I \otimes Z$ for $a = [1,0,1]$) can be computed simply via their bitwise dot product modulo 2 (e.g., $g_1 \cdot a = 1$).

This specific structure allows the output probability $q_\theta(x)$ to be expressed analytically:

\begin{equation} \label{eq:iqp_prob}
q_\theta(x) = 
\frac{1}{2^{2n}} \left| \sum_{z \in \{0,1\}^n} (-1)^{x \cdot z}
\exp\left( -i \sum_{j=1}^m \frac{\theta_j}{2} (-1)^{g_j \cdot z} \right) \right|^2
\end{equation}

\subsection{Classical MMD Training of IQP QCBMs}
While training QCBMs typically requires physical quantum hardware to estimate statistical moments via sampling, recent breakthroughs demonstrate that Pauli-$Z$ expectations for IQP circuits can be evaluated exactly in polynomial time on classical computers \cite{nest2009simulating, recio2025train}. Instead of sampling from the quantum distribution $q_\theta$, the expectation of any parity feature $\langle Z_a \rangle_{q_\theta} = \mathbb{E}_{x \sim q_\theta}[(-1)^{x \cdot a}]$ collapses into a classical expectation over uniformly distributed random bitstrings $z \sim U$:

\begin{equation} \label{eq:model_exp}
\langle Z_a \rangle_{q_\theta} = 
\E[z \sim U]{\cos \left( \sum_{j=1}^m \theta_j (-1)^{g_j \cdot z} (1 - (-1)^{g_j \cdot a}) \right)}
\end{equation}
\emph{Note: since $\theta_j$ is a trainable parameter constant factors such as $\frac{1}{2}$ can be absorbed into it. These do not affect the computation and allow us to keep the notation concise.}

A parameter $\theta_j$ impacts $\langle Z_a \rangle_{q_\theta}$ if and only if $g_j \cdot a = 1 \pmod 2$; otherwise, $(1 - (-1)^{g_j \cdot a}) = 0$.

\textbf{Optimization:}
Leveraging this, \cite{recio2025train} optimized the model classically by minimizing the Maximum Mean Discrepancy (MMD) loss between target and model parity expectations. Using a Gaussian kernel of bandwidth $\sigma$, this loss reformulates exactly into an expectation over observable masks $a$:

\begin{equation} \label{eq:mmd_loss}
\operatorname{MMD}^2(p, q_\theta) = \mathbb{E}_{a \sim P_\sigma}\left[ \left( \langle Z_a \rangle_p - \langle Z_a \rangle_{q_\theta} \right)^2 \right],
\end{equation}
where $P_\sigma(a) = (1 - p_\sigma)^{n - |a|} p_\sigma^{|a|}$ (for $p_\sigma = (1 - e^{-1/2\sigma^2})/2$) acts as a product of independent Bernoulli distributions dependent solely on the Hamming weight $\vert{}a\vert{}$. This enables full classical optimization via Monte Carlo estimation---sampling observables $a \sim P_\sigma$ and bitstrings $z \sim U$---avoiding quantum hardware entirely during training.

\section{Main Results: Gradient Structure and Variance Bounds}
This section will detail the results of this work and interpretations of those results. Derivations for each claim will be provided in the appendices mentioned in each subsection. For clarity of presentation, several substitutions are used in these forms, each of which is listed below.

\begin{equation} \label{eq:substitutions}
\begin{aligned}
    \pi_a &:= \langle Z_a \rangle_p, \\
    f_a(\theta) &:= \langle Z_a \rangle_{q_\theta}, \\
    s_j(z) &:= (-1)^{g_j \cdot z}, \\
    \delta_j(a) &:= 1 - (-1)^{g_j \cdot a}, \\
    \Phi_a(z,\theta) &:= \sum_j \theta_j s_j(z) \delta_j(a).
\end{aligned}
\end{equation}

\subsection{Gradient Sparsity}
\begin{theorem}[Gradient Representation and Sparsity] \label{thm:gradient}
    The gradient of the cost function is as follows:
    \begin{align}
        \pdv{C}{\theta_k} &= 4\sum_{a : g_k \cdot a = 1} P_\sigma(a)(\pi_a - f_a(\theta)) \notag \\
        &\quad \times \E[z\sim U]{s_k(z)\sin(\Phi_a(z,\theta))}
    \end{align}
\end{theorem}

The summation condition in Theorem \ref{thm:gradient}, $ g_k \cdot a = 1$, implies that a parameter only has non-zero gradients for interactions with a restricted subset of observables overlapping with its generator \(g_k\). Thus, sparse IQP generators inherently induce a sparse gradient structure.

The gradient maintains the expectation values of the target dataset and the IQP circuit, while the expectation over the Bernoulli distribution ($P_\sigma(a)$) instead becomes a product by the probability of the given observable from that distribution. The expectation over the uniform distribution remains, with \(s_j(z)\) representing the parity of \(z\) relative to the generator mask \(g_j\) and \(\delta_j(a)\) representing the generator-observable coupling. Specifically, it equals \(2\) if the generator \(g_j\) anticommutes with observable \(Z_a\) or \(0\) if they commute. It serves as a selection mask to determine which circuit generators directly influence the given observable.

As a result of these two components (\(s_j(z)\) and \(\delta_j(a)\)), \(\Phi_a(z,\theta)\) represents the total phase angle accumulated by observable \(Z_a\) across all generators.

\subsection{Lower Bound and Trainability}
The following theorem requires the introduction of a new term, \(\Gamma(a)\), the \emph{reverse light cone} of the observable $Z_a$, defined in Definition~\ref{def:Gamma}.

\begin{definition}[Reverse Light Cone $\Gamma(a)$] \label{def:Gamma}
    For an IQP circuit with generators $g_j$ and an observable $Z_a$, the reverse light cone $\Gamma(a)$ is defined as the set of indices of generators that interact an odd number of times with the observable:
    \[
        \Gamma(a) := \{j : g_j \cdot a = 1\}.
    \]
\end{definition}

To understand the behavior of \(\Gamma(a)\), we briefly derive its meaning in the context of the IQP circuit, then showcase the interactions which occur due to commutation and anti-commutation of operators.

Consider an arbitrary IQP circuit ansatz, \(U = H^{\otimes n}DH^{\otimes n}\) where \(D\) is diagonal in the \(Z\)-basis. Consider an arbitrary \(Z\)-basis observable \(Z_a = \bigotimes_{i\in a}Z_i\) for some subset of qubits \(a\subseteq[n]\). Note that \(H^{\otimes n} Z_a H^{\otimes n} = X_a\) where \(X_a = \bigotimes_{i\in a}X_i\).

Then, the expectation value of the IQP circuit is
\[
\bra{0^{\otimes n}}H^{\otimes n}D^\dagger H^{\otimes n} Z_a H^{\otimes n} D H^{\otimes n} \ket{0^{\otimes n}}.
\]

Simplifying gives
\[
\bra{+^{\otimes n}}D^\dagger X_a D \ket{+^{\otimes n}}.
\]

\begin{figure*}[t!]
    \includegraphics[width=\textwidth]{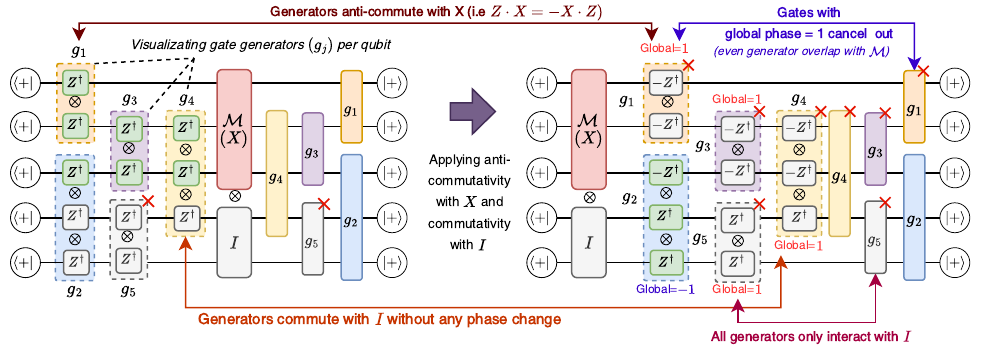}
    \caption{\textbf{Gate commutativity and cancellation in the computation of $\Gamma(a)$}. The figure highlights three types of gates based on their interaction with the observable: (1) Gates with no generator interaction with the observable (e.g., $g_5$) commute and cancel out with their adjoints. (2) Gates with an even number of generator interactions with the observable (e.g., $g_4$) acquire no net phase, allowing them to commute and cancel. (3) Gates with an odd number of generator interactions with the observable (e.g., $g_2$) acquire a negative phase and do not cancel out, forming the effective reverse light cone $\Gamma(a)$.}
    \label{fig:iqp_mpo_parity}
\end{figure*}

This expectation value is visualized using the Matrix Product Operator (MPO) representation as shown in Figure ~\ref{fig:iqp_mpo_parity}. As all the gates in \(D\) are diagonal in the $Z$-basis, we can apply the following simplifications when computing the expectation value.
\begin{enumerate}
    \item Gates whose generators do not overlap with the $X$-basis observable commute with it. This allows them to commute past the observable and other gates, resulting in cancellation with their adjoints. For example the gate resulting from the generator $g_5$ in Figure~\ref{fig:iqp_mpo_parity} cancels out with its adjoint.
    \item Gates whose generators overlap an even number of times with the $X$-basis observable also commute with it and cancel out. Due to the anticommutation of Pauli $X$ and $Z$ operators ($Z_i X_i = -X_i Z_i$), an even number of overlaps between a generator and the observable eliminates any acquired negative signs. For example the gate from the generator $g_4$ in Figure~\ref{fig:iqp_mpo_parity} cancels out with its adjoint. 
\end{enumerate} 

Therefore, the effective reverse light cone with respect to any observable only consists of those gates whose generators overlap with the observable an odd number of times. In other words, \(\Gamma(a)\) is the set of gates that have an effect on the expectation value for a given observable. This work leverages the reverse light cone $\Gamma(a)$ to define a bound under which an arbitrary IQP circuit remains trainable.

% \begin{figure}[b!]
%     \centering
%     \includegraphics[width=0.85\columnwidth]{figures/mpo_full.pdf}
%     \caption{ \textbf{Non-Overlapping Gates:} Gates with generators which do not interact with the observable (ex. $g_2$) can commute with other gates (ex.$g_4$) and combine with their conjugate to cancel out.}
%     \label{fig:iqp_mpo_non_interacting}
% \end{figure}

\begin{theorem}[Gradient Variance Lower Bound] \label{thm:lower_bound}
    The variance of the gradient can be bounded from below by
    \begin{equation}
        \operatorname{Var}\left(\pdv{C}{\theta_k}\right) \geq \sigma_\theta^2\left(\E[\theta]{\pdv{C}{\theta_l}{\theta_k}}\right)^2
    \end{equation}
    for any \(l\). For \(l=k\),
   \begin{equation}
        \begin{aligned}
            \E[\theta]{\pdv[2]{C}{\theta_k}} &= 4 \sum_{a : g_k \cdot a = 1} P_\sigma(a) \delta_k(a) \\
            &\quad \times \bigg[ \pi_a \exp(-2\sigma_\theta^2 |\Gamma(a)|) \E[z\sim U]{\cos(m_1)} \\
            &\qquad - \mathbb{E}_{z,z'\sim U}\Big[ \mathbf{1}(S=-1)\exp(-v_2^-/2)\cos(m_2^-) \\
            &\qquad\qquad\quad + \mathbf{1}(S=1)\exp(-v_2^+/2)\cos(m_2^+) \Big] \bigg]
        \end{aligned}
    \end{equation}

    where
    \begin{align*}
        % \Gamma(a) &:= \{j : g_j \cdot a = 1\}, \\
        m_1 &= \mu_\theta \sum_j s_j(z)\delta_j(a), \\
        m_2^\pm &= \mu_\theta \sum_j \big(s_j(z)\delta_j(a) \pm s_j(z')\delta_j(a)\big), \\
        v_2^+ &= 16\sigma_\theta^2 \sum_{j\in \Gamma(a)}\mathbf{1}(g_j \cdot(z \oplus z') = 0), \\
        v_2^- &= 16\sigma_\theta^2 \sum_{j\in \Gamma(a)}\mathbf{1}(g_j \cdot(z \oplus z') = 1), \\
        S &= \begin{cases}
            1 & g_k \cdot (z \oplus z') = 0, \\
            -1 & g_k \cdot (z \oplus z') = 1.
        \end{cases}
    \end{align*}
    and \(\mathbf{1}(\text{condition})\) is the indicator function.
\end{theorem}

The inequality established in Theorem~\ref{thm:lower_bound} reveals the conditions under which exponential concentration can be avoided, specifically by ensuring that the lower bound does not vanish exponentially. This vanishing can occur through two primary mechanisms:
\begin{enumerate}
    \item \textbf{Destructive Interference:} The expectation value of the Hessian may vanish due to the negative second term. For instance, a highly dense \(\Gamma(a)\) causes \(v_2^\pm\) to approach its mean, which exponentially concentrates the expectation value. The probability of this concentration increases with the size of the reverse light cone, \(|\Gamma(a)|\).
    \item \textbf{Parity Symmetry:} Another potential issue arises from the target distribution \(\pi_a\). If \(\pi_a\) exhibits parity symmetry (e.g., an equal number of zeroes and ones), its expectation value evaluates to zero, which cancels the first term entirely. Nevertheless, provided the second term remains sufficiently large, this cancellation alone does not necessarily induce trainability issues.
\end{enumerate}
% \begin{figure}[t!]
%     \centering
%     \includegraphics[width=0.6\columnwidth]{figures/mpo_final_state.pdf}
%     \caption{Final State remaining after removing non overlapping gates and applying parity simplification.}
%     \label{fig:iqp_mpo_final}
% \end{figure}
% The resulting inequality for Theorem \ref{thm:lower_bound} showcases the conditions under which exponential concentration can be avoided, namely by ensuring the right hand side of the inequality is not exponentially small. First, note that the expectation of the Hessian may vanish as a result of the negative second term. This destructive interference may occur in several cases, including significantly dense \(\Gamma(a)\) which would potentially cause \(v_2^\pm\) to approach their mean and exponentially concentrate the expectation value. The probability of this concentration increases with \(|\Gamma(a)|\). Another potential issue is the target distribution \(\pi_a\). If this distribution has a parity symmetry (e.g., even number of zeroes and ones, then its expectation value would be \(0\), immediately canceling the first term. However, as long as the second term remains sufficiently large, this cancellation would not necessarily induce trainability issues.

Assuming the bound does not vanish as a result of destructive interference, it can be kept non-exponentially small by keeping the term inside the exponential, \(-2\sigma_\theta^2 |\Gamma(a)|\), sufficiently small, causing the exponential function to remain non-exponentially small. The following corollary explores the precise conditions under which this value avoids inducing an exponentially small outcome.

\begin{corollary}[Mitigation of Exponential Concentration] \label{cor:mitigation}
    Assuming destructive interference is avoided as mentioned above, exponential concentration of the gradient variance is mitigated under the following conditions:
    \begin{enumerate}
        \item \textbf{Kernel Bandwidth Scaling:}
        \begin{equation*}
            \sigma = \Omega(n)
        \end{equation*}
        \item \textbf{Initialization Variance Scaling:}
        \begin{equation*}
            \sigma_\theta^2 = O(1/\max_a |\Gamma(a)|)            
        \end{equation*}
        \item \textbf{Parameter Mean Aggregate Rule:}
        \begin{equation*}
            \mu_\theta = O\left(1/{\sqrt{\max_a |\Gamma(a)|}}\right)
        \end{equation*}
        for every observable \(Z_a\).
    \end{enumerate}
\end{corollary}

As highlighted by Corollary \ref{cor:mitigation}, evaluating the lower bound in Theorem~\ref{thm:lower_bound} reveals three key mechanisms that dictate gradient concentration, each suggesting a specific architectural or initialization fix.

\begin{enumerate}
    \item \textbf{Kernel Bandwidth Scaling} A constant kernel bandwidth (\(\sigma = \Theta(1)\)) causes the MMD observable distribution \(P_\sigma(a)\) to concentrate on global observables (\(|a| \sim \Omega(n)\)). These higher-body observables intersect more generators, pushing \(|\Gamma(a)| \to \Omega(n)\) and causing the lower bound to decay exponentially as \(\exp(-\sigma_\theta^2 |\Gamma(a)|)\). Scaling the bandwidth with system size (\(\sigma = \Omega(n)\)) yields \(p_\sigma \approx \Theta(1/n)\), shifting probability mass toward low-body observables (\(|a| \sim O(1)\)) and suppressing light cone growth.
    
    \item \textbf{Initialization Variance vs. Connectivity} Because gradient decay depends on the product \(\sigma_\theta^2 |\Gamma(a)|\), exponential concentration is avoided either by restricting ansatz connectivity to be sparse (\(|\Gamma(a)| = \mathcal{O}(\log n)\)) under constant parameter variance (\(\sigma_\theta^2 = \Omega(1)\)), or by scaling initialization variance inversely with the light-cone size (\(\sigma_\theta^2 = \mathcal{O}(1/|\Gamma(a)|)\)).
    
    \item \textbf{Mean Phase Suppression} A non-zero parameter mean (\(\mu_\theta \neq 0\)) induces destructive phase interference scaling as \(\cos(2\mu_\theta)^{|\Gamma(a)|}\). Setting the parameter mean as \(\mu_\theta = c / \sqrt{|\Gamma(a)|}\) mitigates this decay, yielding \(\lim_{|\Gamma(a)|\to\infty}(1 - 2c^2/|\Gamma(a)|)^{|\Gamma(a)|} = e^{-2c^2}\). This keeps the bound strictly \(\mathcal{O}(1)\) independent of system size. This claim is proved in the proofs section below under the assumption of non-overlapping generators for simplicity. However, overlapping generators may introduce complex interference, which we hypothesize would cause the decay behavior to remain qualitatively similar, as evidenced in part by the mean bound proven in Theorem \ref{thm:lower_bound} and specified in Corollary \ref{cor:mitigation}.
\end{enumerate}

The variance lower bound \(\sigma_\theta^2 = O(1/|\Gamma(a)|)\) establishes a quantum analog to classical Xavier initialization \cite{glorot2010understanding}. Under Xavier initialization in classical deep learning, parameter variance is scaled inversely with fan-in dimension \(d\); here, the size of the reverse light cone \(|\Gamma(a)|\) plays the role of quantum fan-in. Because our approach is purely counting-based, it operates independently of specific geometric graphs, providing substantial flexibility in ansatz design provided \(|\Gamma(a)|\) remains bounded for the target observables. However, our version of the approach requires the variance to be restricted according to the largest \(|\Gamma(a)|\) due to the training methodology randomly sampling observables. Thus, the idea of quantum fan-in is slightly weaker than the classical counterpart.

Our framework also contextualizes recent depth-based asymptotic trainability bounds \cite{shen2026characterizing}, which enforce a narrow initialization variance \(\sigma_\theta^2 = c/D\) (where \(D\) is total parameter count). Theorem~\ref{thm:lower_bound} clarifies why their approach succeeds: since \(|\Gamma(a)| \le D\), substituting their variance yields a dominant factor of \(\exp(-2(c/D)|\Gamma(a)|) \ge \exp(-2c)\), mathematically halting exponential decay. 

However, universally shrinking parameter variance with global depth \(D\) forces parameters near zero, concentrating circuit gates near the identity. As noted by \cite{shen2026characterizing}, this leaves the model vulnerable to efficient classical simulation and severely restricts expressivity. Our light cone framework circumvents this limitation. By pairing a shallow gate architecture (\(|\Gamma(a)| = \mathcal{O}(\log n)\)) with a constant parameter variance (\(\sigma_\theta^2 = \Omega(1)\)), parameter-aware scaling maintains provable trainability while preserving non-trivial state expressivity.

\subsection{Gradient Concentration and Barren Plateaus}
\begin{theorem}[Gradient Concentration Bound] \label{thm:conc_bound}
    The deviation of the gradient from its mean can be probabilistically bounded using a Lipschitz concentration bound for Gaussian random variables:
    \begin{equation}
        \mathbb{P}\left(\left|\pdv{C}{\theta_k} - \E{\pdv{C}{\theta_k}}\right| \geq \epsilon\right) \leq 2\exp(-\frac{\epsilon^2}{2\sigma_\theta^2L^2})
    \end{equation}
    where
    \begin{align*}
        A_{k,l} &:= \{ a\in \{0,1\}^n : g_k \cdot a = 1 \land g_l\cdot a = 1 \}, \\
        L^2 &\leq 576 \sum_l \left(\sum_{a \in A_{k,l}} P_\sigma(a)\right)^2.
    \end{align*}
\end{theorem}

Note that this proof leverages several worst-case bounds and the triangle inequality theorem for computing the Lipschitz constant \(L^2\). As a result, \(L^2\) represents a conservative worst-case bound and the true probabilistic concentration may be tighter in practice.

This concentration bound enables defining the conditions under which exponential concentration is likely to occur. Specifically, the difference between the gradient and its mean can be shown not to exceed an exponentially small \(\epsilon\) with high probability by minimizing the exponential function, which occurs when \(\sigma_\theta^2 L^2\) is kept sufficiently small.

\begin{corollary}[Source of Exponential Concentration]
    For a fixed deviation threshold \(\epsilon\), the probability of observing a deviation higher than \(\epsilon\) decays exponentially when \(\sigma_\theta^2L^2\) is sufficiently small. To ensure that the probability of a deviation exceeding \(\epsilon\) is at most \(\delta\), it suffices that
    \begin{equation}
        \sigma_\theta^2L^2 \leq \frac{\epsilon^2}{2\ln\left(\frac{2}{\delta}\right)}.
    \end{equation}
\end{corollary}

Due to the interaction between \(\sigma_\theta^2\) and \(L^2\), it is possible that sufficiently reducing one quantity and increasing the other may balance the two and not necessarily result in an exponential concentration. Thus, the two cases to consider are small \(\sigma_\theta^2\), which is set in advance, or small \(L^2\).

The magnitude of the Lipschitz constant \(L^2\) is controlled by the interaction between the generator overlap set \(A_{k,l}\) and the distribution \(P_\sigma(a)\). Specifically, it depends on two factors: the number of active terms in the outer sum, \(\sum_l\), and the probability mass assigned by the inner sum, \(\sum_{a\in A_{k,l}}P_\sigma(a)\).

When the ansatz architecture is sparse, individual generators only act on a small number of qubits. Consequently, a given generator \(g_k\) shares overlapping anti-commuting observables (i.e., \(A_{k,l} \neq \emptyset\)) with very few other generators \(g_l\). This significantly restricts the outer sum, \(\sum_l\), to a small number of non-zero terms, keeping \(L^2\) bounded. However, the observables in \(A_{k,l}\) would have low Hamming weight. When the target bandwidth \(\sigma\) is small, \(P_\sigma(a)\) concentrates its probability mass onto these low-body observables, reducing the probability of exponential concentration.

In contrast, a dense ansatz architecture creates extensive overlap among generators, causing the outer sum to contain significantly more non-zero terms. However, the observables required to satisfy \(a\in A_{k,l}\) generally possess higher Hamming weight. When the target bandwidth \(\sigma\) is large, \(P_\sigma(a)\) penalizes high-body observables, suppressing the inner sum. A smaller bandwidth similarly results in a small inner sum due to the exponential number of high-body observables. In both cases, this small inner sum offsets the larger number of terms in the outer sum, pushing \(L^2\) towards zero.

% The second condition is controlled by the summation target \(A_{k,l}\). If there is little overlap between generators and observables (e.g., if the ansatz has a sparse architecture), then the outer sum of \(L^2\), i.e., \(\sum_l\), will likely contain very few non-zero terms. This outer sum would then be restricted to very few terms. The inner sum depends in part on the bandwidth parameter, \(\sigma^2\). Since the ansatz is sparse, the observables that align with the generators will similarly be low-bodied and therefore have a low Hamming weight, \(|a| \sim O(1)\). If the bandwidth is sufficiently small, then \(P_\sigma(a)\) will be exponentially small, resulting in a decreasing value for \(L^2\). If \(\sigma^2 \sim \Omega(n)\), then \(P_\sigma(a)\) will only be polynomially small. However, given the restriction of the outer sum, this summation will be limited in either case for a sparse ansatz architecture.

% For a denser ansatz architecture, the outer sum will have significantly more non-zero terms, potentially \(\Omega(n)\) or \(\Omega(n^2)\) depending on the specific circuit architecture. The main issue in this case is the inner sum. The observables which align with a denser architecture will have higher Hamming weight, \(|a| \sim \Omega(n)\). Regardless of choice of bandwidth, the probability \(P_\sigma(a)\) will be exponentially small.

In cases where the gradient is exponentially concentrated, barren plateaus occur in the case where the mean of the gradient is \(0\). The mean of the gradient is as follows.

\begin{align*}
    \E[\theta]{\pdv{C}{\theta_k}} 
    &= \E[\theta]{4\sum_{a : g_k \cdot a = 1} P_\sigma(a)(\pi_a - f_a(\theta))X_a} \\
    &= 4\sum_{a : g_k \cdot a = 1} P_\sigma(a)\big(\pi_a \E[\theta]{X_a} - \E[\theta]{f_aX_a}\big) \\[0.4em]
    \E[\theta]{X_a} 
    &= \E[z]{s_k(z)\E[\theta]{\sin(\Phi_a(z,\theta))}} \\[0.4em]
    \E[\theta]{f_aX_a} 
    &= \E[z,z']{s_k(z)\E[\theta]{\cos(\Phi_a(z', \theta))\sin(\Phi_a(z,\theta))}} \\
    &= \frac{1}{2}\mathbb{E}_{z,z'}\Big[ s_k(z)\mathbb{E}_\theta\big[ \sin(\Phi_a(z, \theta) + \Phi_a(z',\theta)) \\
    &\qquad\qquad\qquad + \sin(\Phi_a(z, \theta) - \Phi_a(z',\theta)) \big] \Big]
\end{align*}

Each of these expectation values with respect to \(\theta\) depend multiplicatively on \(\sin(m)\) where \(m\) depends multiplicatively on \(\mu_\theta\) as discussed in the proof of Theorem \ref{thm:lower_bound}. Thus, the mean of the gradient is \(0\) when the mean of \(\theta\) is \(0\). However, barren plateaus require exponential concentration of the variance of the gradient in addition to the mean concentrating at \(0\). Thus, a small mean does not guarantee barren plateaus on its own, and may remain an effective strategy in mitigating exponential concentration.

\section{Conclusion and Future Work}
This work provided an analytical lower bound on the variance of the gradient and a probabilistic concentration bound on the deviation of the gradient from its mean, enabling analysis of the factors under which exponential concentration is more or less likely to occur. While mitigating exponential concentration can support trainability, it does not guarantee effective training of the model or good generalization performance. Notably, the strategies discussed in this work are data independent and do not take into consideration any inductive bias that the ansatz may offer. 
% For example, a circuit which is well aligned with the data may not suffer as readily from exponential concentration despite poor choices for parameter mean or kernel bandwidth due to alignment of the generators with the data in such a way that gradients are preserved.

Future work may explore some of these ideas experimentally, such as exploring the interaction between a small bandwidth and a circuit architecture that is either aligned to or random with respect to the data. Various strategies have been offered to mitigate barren plateaus in these IQP QCBM models using uniform sampling from certain distributions. Future work may also explore whether those strategies prove effective in the case where parameters are sampled from a Gaussian distribution.

\printbibliography

\newpage
\appendix
\section{Proofs}
\subsection{Proof of Theorem 1 - Gradient Representation and Sparsity} \label{app:derivatives}
We start by defining the expected values of the Pauli-$Z$ observable $a$ for the target distribution $p$ and the parameterized model distribution $q_\theta$:
\begin{align*}
    \pi_a       & := \langle Z_a \rangle_p          \\
    f_a(\theta) & := \langle Z_a \rangle_{q_\theta}
\end{align*}

The cost function $C(\theta)$ is defined using the Maximum Mean Discrepancy (MMD) between $p$ and $q_\theta$, weighted by a prior distribution $P_\sigma$ over the observables:
\begin{align*}
    C(\theta) & = \operatorname{MMD}(p, q_\theta) = \E[a \sim P_\sigma]{(\pi_a - f_a(\theta))^2} \\
              & = \sum_a P_\sigma(a)(\pi_a - f_a(\theta))^2
\end{align*}

For an IQP Quantum Circuit Born Machine, the output distribution's expectation value can be written analytically over the uniformly distributed boolean variables $z$:
\begin{align*}
    f_a(\theta) & = \E[z \sim U]{\cos \left( \sum_j \theta_j (-1)^{g_j \cdot z}(1-(-1)^{g_j \cdot a}) \right)}
\end{align*}

To simplify the expression, we define the sign variable $s_j(z)$ and the observable overlap indicator $\delta_j(a)$:
\begin{align*}
    s_j(z)      & := (-1)^{g_j \cdot z}     \\
    \delta_j(a) & := 1 - (-1)^{g_j \cdot a}
\end{align*}

Substituting these definitions, the expectation value becomes:
\begin{align*}
    f_a(\theta) & = \E[z \sim U]{\cos \left( \sum_j \theta_j s_j(z) \delta_j(a) \right)}
\end{align*}

We can further compress this by defining the phase $\Phi_a(z, \theta)$:
\begin{align*}
    \Phi_a(z, \theta) & := \sum_j \theta_j s_j(z) \delta_j(a) \\
    f_a(\theta)       & = \E[z \sim U]{\cos(\Phi_a)}
\end{align*}

Now, we take the derivative of the cost function with respect to a parameter $\theta_k$. Applying the chain rule to the MMD cost yields:
\begin{align*}
    \pdv{C}{\theta_k} & = \sum_a P_\sigma(a) \cdot 2(\pi_a - f_a(\theta))\left(-\pdv{f_a(\theta)}{\theta_k}\right) \\
                      & = -2 \sum_a P_\sigma(a)(\pi_a - f_a(\theta))\pdv{f_a(\theta)}{\theta_k}
\end{align*}

Next, we evaluate the derivative of $f_a(\theta)$ with respect to $\theta_k$:
\begin{align*}
    \pdv{f_a(\theta)}{\theta_k} & = \E[z \sim U]{-\sin(\Phi_a) \pdv{\Phi_a}{\theta_k}}
\end{align*}

Since $\Phi_a$ is linear in the parameters $\theta_j$, its partial derivative is simply the coefficient of $\theta_k$:
\begin{align*}
    \pdv{\Phi_a}{\theta_k} & = s_k(z)\delta_k(a)
\end{align*}

Thus, the gradient of the expectation value is:
\begin{align*}
    \pdv{f_a(\theta)}{\theta_k} & = -\E[z \sim U]{s_k(z)\delta_k(a)\sin(\Phi_a)}
\end{align*}

Plugging this back into the derivative of the cost function, we obtain the full gradient:
\begin{align*}
    \pdv{C}{\theta_k} & = -2 \sum_a P_\sigma(a)(\pi_a - f_a(\theta))\left( -\E[z \sim U]{s_k(z)\delta_k(a)\sin(\Phi_a)} \right) \\
                      & = 2 \sum_a P_\sigma(a)(\pi_a - f_a)\cdot\E[z \sim U]{s_k(z) \delta_k(a) \sin(\Phi_a(z, \theta))}
\end{align*}

We can simplify the overlap indicator $\delta_k(a)$ by evaluating it over the binary inner product:
\begin{align*}
    \delta_k(a) & = 1 - (-1)^{g_k \cdot a}
\end{align*}

Since $(-1)^{g_k \cdot a} \in \{ \pm 1 \}$ and the inner product $g_k \cdot a \in \mathbb{F}_2$, we have:
\begin{align*}
    \delta_k(a) & =
    \begin{cases}
        0 & g_k \cdot a = 0 \\
        2 & g_k \cdot a = 1
    \end{cases}
\end{align*}

Therefore, only observables $a$ that overlap with a generator $g_k$ (i.e., $g_k \cdot a = 1$) will contribute a non-zero term to the gradient. Applying this sparsity condition gives the final simplified gradient:
\begin{align*}
    \pdv{C}{\theta_k} & = 4 \sum_{a : g_k \cdot a = 1} P_\sigma(a)(\pi_a - f_a)\cdot\E[z \sim U]{s_k(z) \sin(\Phi_a(z, \theta))}
\end{align*}

\subsection{Proof of Theorem 2 - Gradient Variance Lower Bound} \label{app:lower_bound}
To establish the variance bound, we employ the covariance inequality:
\begin{align*}
    \operatorname{Cov}(X, Y)^2 &\leq \operatorname{Var}(X)\operatorname{Var}(Y)
\end{align*}

Rearranging this inequality provides a lower bound on the variance of $X$:
\begin{align*}
    \operatorname{Var}(X) &\geq \frac{\operatorname{Cov}(X, Y)^2}{\operatorname{Var}(Y)}
\end{align*}

We apply this to the gradient components, letting $X$ be the gradient with respect to $\theta_k$ and $Y$ be another parameter $\theta_l$. We assume the parameters $\theta$ are initialized from a Gaussian distribution:
\begin{align*}
    X &= \pdv{C}{\theta_k}, \quad Y = \theta_l \\
    \theta &\sim \mathcal{N}(\mu_\theta,\sigma_\theta^2)
\end{align*}

This allows us to bound the variance of the gradient component:
\begin{align*}
    \operatorname{Var}\left(\pdv{C}{\theta_k}\right) &\geq \frac{\operatorname{Cov}\left(\pdv{C}{\theta_k}, \theta_l\right)^2}{\operatorname{Var}(\theta_l)} \\
    &= \frac{\operatorname{Cov}\left(\pdv{C}{\theta_k}, \theta_l\right)^2}{\sigma_\theta^2}
\end{align*}

Next, we prove an identity for the covariance of a function of a Gaussian variable with the variable itself. By definition:
\begin{align*}
    \operatorname{Cov}(A,B) &= \E{(A - \E{A})(B - \E{B})}
\end{align*}

Let $Y \sim \mathcal{N}(\mu, \sigma^2)$ be a Gaussian random variable. The covariance between $f(Y)$ and $Y$ is:
\begin{align*}
    \operatorname{Cov}(f(Y),Y) &= \E{(f(Y) - \E{f(Y)})(Y - \E{Y})} \\
    &= \E{(f(Y) - \E{f(Y)})(Y - \mu)} \\
    &= \E{f(Y)(Y - \mu)} - \E{\E{f(Y)}(Y - \mu)}
\end{align*}

Since the expectation of $f(Y)$ is a constant, we can factor it out:
\begin{align*}
    \operatorname{Cov}(f(Y),Y) &= \E{f(Y)(Y - \mu)} - \E{f(Y)}\E{Y - \mu}
\end{align*}

The expectation $\E{Y - \mu}$ evaluates to zero:
\begin{align*}
    \E{Y - \mu} &= \E{Y} - \mu = \mu - \mu = 0
\end{align*}

Thus, the covariance simplifies to:
\begin{align*}
    \operatorname{Cov}(f(Y),Y) &= \E{f(Y)(Y - \mu)}
\end{align*}

By applying Stein's lemma for Gaussian variables, we can further rewrite this covariance in terms of the expected derivative:
\begin{align*}
    \operatorname{Cov}(f(Y),Y) &= \sigma^2\E{f'(Y)}
\end{align*}

For multivariate functions with independent and identically distributed (i.i.d.) parameters, this generalizes to:
\begin{align*}
    \E{f(Y)(Y_l - \mu)} &= \sigma^2\E{\pdv{f(Y)}{Y_l}}
\end{align*}

Returning to the original variance lower bound, we apply Stein's lemma where $f(Y)$ is the gradient $\pdv{C}{\theta_k}$ and $Y_l$ is the parameter $\theta_l$:
\begin{align*}
    \operatorname{Var}\left(\pdv{C}{\theta_k}\right) &\geq \frac{\operatorname{Cov}\left(\pdv{C}{\theta_k}, \theta_l\right)^2}{\sigma_\theta^2} \\
    &= \frac{\left(\sigma_\theta^2 \E[\theta]{\pdv{C}{\theta_l}{\theta_k}}\right)^2}{\sigma_\theta^2} \\
    &= \sigma_\theta^2\left(\E[\theta]{\pdv{C}{\theta_l}{\theta_k}}\right)^2
\end{align*}

Showing that this expectation does not exponentially decay for even a single value of $l$ is sufficient to prove that the overall variance of the gradient does not exponentially decay (i.e., there is no barren plateau). From the previous section, the gradient is given by:
\begin{align*}
    \pdv{C}{\theta_k} &= 4\sum_{a : g_k \cdot a = 1}P_\sigma(a)(\pi_a - f_a(\theta))X_a
\end{align*}

Where we define the auxiliary variable $X_a$ and recall the previous definitions:
\begin{align*}
    X_a &:= \E[z \sim U]{s_k(z)\sin(\Phi_a(z,\theta))} \\
    \Phi_a(z,\theta) &:= \sum_j \theta_j s_j(z) \delta_j(a) \\
    s_k(z) &:= (-1)^{g_k \cdot z} \\
    \delta_j(a) &:= 1 - (-1)^{g_j \cdot a} \\
    f_a(\theta) &:= \E[z \sim U]{\cos(\Phi_a(z,\theta))}
\end{align*}

Taking the mixed partial derivative with respect to $\theta_l$ and applying the product rule yields:
\begin{align*}
    \pdv{C}{\theta_l}{\theta_k} &= 4\sum_{a : g_k \cdot a = 1} P_\sigma(a)\left[ (\pi_a - f_a(\theta))\pdv{X_a}{\theta_l} + \left(-\pdv{f_a}{\theta_l}(\theta)\right)X_a \right]
\end{align*}

Where the internal partial derivatives are evaluated as:
\begin{align*}
    \pdv{X_a}{\theta_l} &= \E[z]{s_k(z)s_l(z)\delta_l(a)\cos(\Phi_a(z,\theta))} \\
    \pdv{f_a}{\theta_l} &= -\E[z]{s_l(z)\delta_l(a)\sin(\Phi_a(z,\theta))}
\end{align*}

To evaluate expectations of these trigonometric functions over the Gaussian parameters $\theta$, we first need a general method to compute the expected sine and cosine of a Gaussian random variable $X \sim \mathcal{N}(m,v)$. Using the characteristic function of the Gaussian distribution:
\begin{align*}
    \E{e^{itX}} &= e^{itm-\frac{1}{2}t^2v} \\
    \E{e^{iX}} &= e^{im-v/2}
\end{align*}

By expanding the complex exponential using Euler's formula:
\begin{align*}
    e^{iX} &= \cos(X)+i\sin(X) \\
    \E{e^{iX}} &= \E{\cos(X)} + i\E{\sin(X)}
\end{align*}

Equating this to the characteristic function evaluated at $t=1$:
\begin{align*}
    \E{e^{iX}} &= e^{im-v/2} \\
    &= e^{-v/2}e^{im}
\end{align*}

And applying Euler's formula again to $e^{im}$:
\begin{align*}
    e^{im} &= \cos(m) + i\sin(m) \\
    \E{e^{iX}} &= e^{-v/2}(\cos(m) + i\sin(m)) \\
    &= e^{-v/2}\cos(m) + ie^{-v/2}\sin(m)
\end{align*}

Matching the real and imaginary parts of these two expansions gives the exact expectations:
\begin{align*}
    \E{\cos(X)} &= e^{-v/2}\cos(m) \\
    \E{\sin(X)} &= e^{-v/2}\sin(m)
\end{align*}

We will need the mean and variance of linear combinations of Gaussian variables, such as $\Phi_a$ or $\Phi_a \pm \Phi_b$. Consider a general linear combination where the coefficients $K_j$ are independent of $\theta$:
\begin{align*}
    X &= \sum_j \theta_j K_j
\end{align*}

Its mean and variance are simply:
\begin{align*}
    \E[\theta]{X} &= \sum_j K_j \E[\theta]{\theta_j} = \mu_\theta \sum_j K_j \\
    \operatorname{Var}(X) &= \operatorname{Var}\left( \sum_j K_j \theta_j \right) \\
    &= \sum_j K_j^2 \operatorname{Var}(\theta_j) = \sigma_\theta^2 \sum_j K_j^2
\end{align*}

In our derivation, we encounter two main cases for the coefficients $K_j$. Case 1 involves the single phase variable:
\begin{align*}
    X &= \Phi_a(z, \theta) = \sum_j \theta_j s_j(z) \delta_j(a) \\
    K_j &= s_j(z)\delta_j(a)
\end{align*}

Case 2 involves the sum or difference of two phase variables at independent boolean inputs $z$ and $z'$:
\begin{align*}
     X &= \Phi_a(z, \theta) \pm \Phi_a(z',\theta) \\
     &= \sum_j \theta_j s_j(z) \delta_j(a) \pm \sum_j \theta_j s_j(z') \delta_j(a) \\
     &= \sum_j \theta_j (s_j(z)\delta_j(a) \pm s_j(z')\delta_j(a)) \\
     K_j &= s_j(z)\delta_j(a) \pm s_j(z')\delta_j(a)
\end{align*}

Returning to the mixed partial derivative (the Hessian element) and taking its expectation over the parameters $\theta$:
\begin{align*}
    \E[\theta]{\pdv{C}{\theta_l}{\theta_k}} &= 4\sum_{a : g_k \cdot a = 1} P_\sigma(a)\left[ \pi_a\E[\theta]{\pdv{X_a}{\theta_l}} - \E[\theta]{f_a \pdv{X_a}{\theta_l}} + \E[\theta]{\left(-\pdv{f_a}{\theta_l}\right) X_a} \right]
\end{align*}

We can analyze each of the three terms in the brackets separately. Let's begin with the first expectation:
\begin{align*}
    \E[\theta]{\pdv{X_a}{\theta_l}} &= \E[\theta]{\E[z]{s_k(z)s_l(z)\delta_l(a)\cos(\Phi_a(z, \theta))}} \\
    &= \E[z]{s_k(z)s_l(z)\delta_l(a)\E[\theta]{\cos(\Phi_a(z, \theta))}}
\end{align*}

Using our formula for the expectation of a cosine of a Gaussian, with mean $m_1$ and variance $v_1$:
\begin{align*}
    \E[\theta]{\cos(\Phi_a)} &= e^{-v_1/2}\cos(m_1) \\
    m_1 &= \mu_\theta \sum_j s_j(z)\delta_j(a) \\
    v_1 &= \sigma_\theta^2 \sum_j s_j(z)^2\delta_j(a)^2
\end{align*}

Recall the definition of the indicator $\delta_j(a)$:
\begin{align*}
    \delta_j(a) &= 
    \begin{cases}
        0 & g_j \cdot a = 0 \\
        2 & g_j \cdot a = 1
    \end{cases}
\end{align*}

Defining $\Gamma(a)$ as the set of indices where the observable overlaps with a generator:
\begin{align*}
    \Gamma(a) &:= \{j : g_j \cdot a = 1\}
\end{align*}

Since $s_j(z) \in \{\pm 1\}$, its square is $s_j(z)^2 = 1$. This allows us to simplify the variance $v_1$:
\begin{align*}
    v_1 &= 4\sigma_\theta^2 \sum_{j \in \Gamma(a)}1 = 4\sigma_\theta^2 |\Gamma(a)|
\end{align*}

Plugging this back into the expected cosine:
\begin{align*}
    \E[\theta]{\cos(\Phi_a)} &= \exp(-2\sigma_\theta^2 |\Gamma(a)|)\cos\left(\mu_\theta \sum_j s_j(z)\delta_j(a)\right)
\end{align*}

This completes the first term of the Hessian expectation:
\begin{align*}
    \E[\theta]{\pdv{X_a}{\theta_l}} &= \E[z]{s_k(z)s_l(z)\delta_l(a)\E[\theta]{\cos(\Phi_a(z, \theta))}} \\
    &= \E[z]{s_k(z)s_l(z)\delta_l(a)\exp(-2\sigma_\theta^2 |\Gamma(a)|)\cos\left(\mu_\theta \sum_j s_j(z)\delta_j(a)\right)}
\end{align*}

Now we evaluate the second term, which involves a product of expectations over the independent boolean variables $z$ and $z'$:
\begin{align*}
    \E[\theta]{f_a \pdv{X_a}{\theta_l}} &= \E[\theta]{\E[z]{\cos(\Phi_a(z,\theta))}\E[z']{s_k(z')s_l(z')\delta_l(a)\cos(\Phi_a(z',\theta))}} \\
    &= \E[\theta]{\E[z,z']{s_k(z')s_l(z')\delta_l(a)\cos(\Phi_a(z,\theta))\cos(\Phi_a(z',\theta))}} \\
    &= \E[z,z']{s_k(z')s_l(z')\delta_l(a)\E[\theta]{\cos(\Phi_a(z,\theta))\cos(\Phi_a(z',\theta))}}
\end{align*}

To compute this, we use the trigonometric product-to-sum identity. Letting $\Phi_a := \Phi_a(z,\theta)$ and $\Phi_a' := \Phi_a(z', \theta)$:
\begin{align*}
    \E[\theta]{\cos(\Phi_a)\cos(\Phi_a')} &= \frac{1}{2}\E[\theta]{\cos(\Phi_a - \Phi_a') + \cos(\Phi_a + \Phi_a')}
\end{align*}

We again apply our expectation formula for Gaussians, where $m_2^\pm$ and $v_2^\pm$ are the mean and variance of $\Phi_a \pm \Phi_a'$:
\begin{align*}
    \E[\theta]{\cos(\Phi_a \pm \Phi_a')} &= e^{-v_2^\pm/2}\cos(m_2^\pm)
\end{align*}

Where the mean and variance are given by:
\begin{align*}
    m_2^\pm &= \mu_\theta \sum_j \left(s_j(z)\delta_j(a) \pm s_j(z')\delta_j(a)\right) \\
    v_2^\pm &= \sigma_\theta^2 \sum_j (s_j(z)\delta_j(a) \pm s_j(z')\delta_j(a))^2 \\
    &= \sigma_\theta^2 \sum_j \left(s_j(z)^2\delta_j(a)^2 + s_j(z')^2\delta_j(a)^2 \pm 2s_j(z)s_j(z')\delta_j(a)^2\right) \\
    &= 4\sigma_\theta^2\left(\sum_{j \in \Gamma(a)} (1 + 1 \pm 2s_j(z)s_j(z'))\right) \\
    &= 8\sigma_\theta^2\left(\sum_{j \in \Gamma(a)} (1 \pm s_j(z)s_j(z'))\right)
\end{align*}

Notice that the product $s_j(z)s_j(z')$ evaluates to $s_j(z \oplus z')$. We analyze the two possible values of the generator inner product:
\begin{align*}
    \text{If } g_j \cdot (z \oplus z') = 1 &\Rightarrow (-1)^{g_j\cdot(z \oplus z')} = -1 \\
    \text{If } g_j \cdot (z \oplus z') = 0 &\Rightarrow (-1)^{g_j\cdot(z \oplus z')} = 1
\end{align*}

Using indicator functions, the variances can be written succinctly as:
\begin{align*}
    v_2^+ &= 16\sigma_\theta^2 \sum_{j\in \Gamma(a)}\mathbf{1}(g_j \cdot(z \oplus z') = 0) \\
    v_2^- &= 16\sigma_\theta^2 \sum_{j\in \Gamma(a)}\mathbf{1}(g_j \cdot(z \oplus z') = 1)
\end{align*}

Bringing this back, the expected product of cosines is:
\begin{align*}
    \E[\theta]{\cos(\Phi_a)\cos(\Phi_a')} &= \frac{1}{2}\left(e^{-v_2^-/2}\cos(m_2^-) + e^{-v_2^+/2}\cos(m_2^+)\right)
\end{align*}

And the full second term of the Hessian expectation evaluates to:
\begin{align*}
    \E[\theta]{f_a \pdv{X_a}{\theta_l}} &= \frac{1}{2}\E[z,z']{s_k(z')s_l(z')\delta_l(a)\left(e^{-v_2^-/2}\cos(m_2^-) + e^{-v_2^+/2}\cos(m_2^+)\right)}
\end{align*}

Now we move to the third and final term, which involves a product of sines:
\begin{align*}
    \E[\theta]{\left(-\pdv{f_a}{\theta_l}\right) X_a} &= \E[\theta]{\E[z,z']{s_l(z)\delta_l(a)\sin(\Phi_a)s_k(z')\sin(\Phi_a')}} \\
    &= \E[z,z']{s_l(z)s_k(z')\delta_l(a)\E[\theta]{\sin(\Phi_a)\sin(\Phi_a')}}
\end{align*}

Applying the corresponding trigonometric product-to-sum identity for sines:
\begin{align*}
     \E[\theta]{\sin(\Phi_a)\sin(\Phi_a')} &= \frac{1}{2}\E[\theta]{\cos(\Phi_a - \Phi_a') - \cos(\Phi_a + \Phi_a')}
\end{align*}

Since this involves the same arguments $\Phi_a \pm \Phi_a'$, the means and variances are identical to those computed for the second term:
\begin{align*}
     \E[\theta]{\sin(\Phi_a)\sin(\Phi_a')} &= \frac{1}{2}\left(e^{-v_2^-/2}\cos(m_2^-) - e^{-v_2^+/2}\cos(m_2^+)\right)
\end{align*}

Giving the final evaluation of the third term:
\begin{align*}
     \E[\theta]{\left(-\pdv{f_a}{\theta_l}\right) X_a} &= \frac{1}{2}\E[z,z']{s_l(z)s_k(z')\delta_l(a)\left(e^{-v_2^-/2}\cos(m_2^-) - e^{-v_2^+/2}\cos(m_2^+)\right)}
\end{align*}

To recap, we have expanded the expected Hessian element into three distinct components:
\begin{align*}
    \E[\theta]{\pdv{C}{\theta_l}{\theta_k}} &= 4\sum_{a : g_k \cdot a = 1} P_\sigma(a)\left[ \pi_a\E[\theta]{\pdv{X_a}{\theta_l}} - \E[\theta]{f_a \pdv{X_a}{\theta_l}} + \E[\theta]{\left(-\pdv{f_a}{\theta_l}\right) X_a} \right]
\end{align*}

To show that the variance does not vanish, it is sufficient to evaluate this bound at $l=k$, which simplifies the expressions considerably since $s_k(z)^2 = 1$:
\begin{align*}
    \E[\theta]{\pdv{X_a}{\theta_k}} &= \E[z]{\delta_k(a)\exp(-2\sigma_\theta^2 |\Gamma(a)|)\cos\left(\mu_\theta \sum_j s_j(z)\delta_j(a)\right)} \\
    \E[\theta]{f_a \pdv{X_a}{\theta_k}} &= \frac{1}{2}\E[z,z']{\delta_k(a)\left(e^{-v_2^-/2}\cos(m_2^-) + e^{-v_2^+/2}\cos(m_2^+)\right)} \\
    \E[\theta]{\left(-\pdv{f_a}{\theta_k}\right) X_a} &= \frac{1}{2}\E[z,z']{s_k(z \oplus z')\delta_k(a)\left(e^{-v_2^-/2}\cos(m_2^-) - e^{-v_2^+/2}\cos(m_2^+)\right)}
\end{align*}

We can combine the second and third terms by factoring out common elements. Let $S := s_k(z \oplus z')$, $A := e^{-v_2^-/2}\cos(m_2^-)$, and $B := e^{-v_2^+/2}\cos(m_2^+)$:
\begin{align*}
    -\frac{1}{2}(A + B) + \frac{1}{2}S(A - B) &= \frac{1}{2}[A(S - 1) - B(S + 1)]
\end{align*}

The sign variable $S$ can only take two values, which act as a selector between $A$ and $B$:
\begin{align*}
    S &=
    \begin{cases}
        1 & g_k \cdot (z \oplus z') = 0 \\
        -1 & g_k \cdot (z \oplus z') = 1
    \end{cases} \\
    S = 1 &\Rightarrow \frac{1}{2}[A(1 - 1) - B(1 + 1)] = -B \\
    S = -1 &\Rightarrow \frac{1}{2}[A(-1 - 1) - B(-1 + 1)] = -A
\end{align*}

Bringing everything together, the full expectation of the Hessian is:
\begin{align*}
    \E[\theta]{\pdv[2]{C}{\theta_k}} &= 4 \sum_{a : g_k \cdot a = 1} P_\sigma(a) \delta_k(a) \Bigg[ \pi_a \E[z]{\exp(-2\sigma_\theta^2 |\Gamma(a)|)\cos(m_1)} \\
    &\quad - \E[z,z']{\mathbf{1}(S=-1)e^{-v_2^-/2}\cos(m_2^-) + \mathbf{1}(S=1)e^{-v_2^+/2}\cos(m_2^+)} \Bigg]
\end{align*}

\subsection{Proof of Theorem 3 - Gradient Concentration Bound} \label{app:upper_bound}
Using the Hessian derived above, it is also possible to derive a probabilistic concentration bound using Lipschitz concentration for Gaussian random variables. To start, this section will show that the gradient \(\pdv{C}{\theta_k}\) is an \(L\)-Lipschitz function of \(\theta\). This is shown by bounding the \(L_2\) norm of its gradient vector, whose \(l\)-th component is the mixed partial derivative \(\pdv{C}{\theta_l}{\theta_k}\), by \(L\) for any input \(\theta\). We formalize this bound as:
\begin{align*}
    L^2 &= \sup_\theta \sum_l \left(\pdv{C}{\theta_l}{\theta_k}\right)^2
\end{align*}

From our previous derivations, the elements of the Hessian can be expanded as follows:
\begin{align*}
    \pdv{C}{\theta_l}{\theta_k} &= 4\sum_{a : g_k \cdot a = 1} P_\sigma(a)\left[ (\pi_a - f_a(\theta))\pdv{X_a}{\theta_l} + \left(-\pdv{f_a}{\theta_l}(\theta)\right)X_a \right]
\end{align*}

Where the partial derivatives of the internal components $X_a$ and $f_a$ with respect to $\theta_l$ are given by:
\begin{align*}
    \pdv{X_a}{\theta_l} &= \E[z]{s_k(z)s_l(z)\delta_l(a)\cos(\Phi_a(z,\theta))} \\
    \pdv{f_a}{\theta_l} &= -\E[z]{s_l(z)\delta_l(a)\sin(\Phi_a(z,\theta))}
\end{align*}

We start by maximizing each of these components individually to find a worst-case upper bound. The target probability $\pi_a$ and expected model output $f_a(\theta)$ both represent expectations of Pauli observables, thus they are bounded:
\begin{align*}
    \pi_a, f_a(\theta) &\in [-1, 1] \\
    |\pi_a - f_a(\theta)| &\leq 2
\end{align*}

Similarly, the basis functions are bounded by unity:
\begin{align*}
    |s_l(z)| &= 1 \\
    \sin(\cdot), \cos(\cdot) &\in [-1,1] \\
    |X_a| &\leq 1
\end{align*}

Applying these bounds to the partial derivatives, we can bound their absolute values by the overlap indicator $\delta_l(a)$:
\begin{align*}
    \left|\pdv{f_a}{\theta_l}(\theta)\right| &\leq \delta_l(a) \\
    \left|\pdv{X_a}{\theta_l}\right| &\leq \delta_l(a)
\end{align*}

Combining all these bounds back into the absolute value of the Hessian component:
\begin{align*}
    \left|\pdv{C}{\theta_l}{\theta_k}\right| &\leq 4\sum_{a : g_k \cdot a = 1} P_\sigma(a)\left[ 2\delta_l(a) + \delta_l(a) \right] \\
    &= 12\sum_{a : g_k \cdot a = 1} P_\sigma(a)\delta_l(a)
\end{align*}

To further simplify this, we define the set of observables $A_{k,l}$ that simultaneously overlap with both generators $g_k$ and $g_l$:
\begin{align*}
    A_{k,l} &:= \{ a\in \{0,1\}^n : g_k \cdot a = 1 \land g_l\cdot a = 1 \}
\end{align*}

Since $\delta_l(a)$ evaluates to 2 when $g_l \cdot a = 1$ and 0 otherwise:
\begin{align*}
    \delta_l(a) &=
    \begin{cases}
        2 & g_l \cdot a = 1 \\
        0 & g_l \cdot a = 0
    \end{cases}
\end{align*}

We can rewrite the sum over only the observables in the intersection set $A_{k,l}$:
\begin{align*}
    \left|\pdv{C}{\theta_l}{\theta_k}\right| &\leq 24\sum_{a \in A_{k,l}} P_\sigma(a)
\end{align*}

Squaring this bound and summing over all parameters $l$ gives us the final Lipschitz constant $L^2$:
\begin{align*}
    L^2 &\leq \sum_l \left( 24\sum_{a \in A_{k,l}} P_\sigma(a) \right)^2 \\
    &= 576 \sum_l \left(\sum_{a \in A_{k,l}} P_\sigma(a)\right)^2
\end{align*}

Finally, applying the standard Gaussian concentration bound for Lipschitz functions, we can bound the probability that the gradient deviates from its expected value:
\begin{align*}
    \mathbb{P}\left(\left|\pdv{C}{\theta_k} - \E{\pdv{C}{\theta_k}}\right| \geq \epsilon\right) &\leq 2\exp\left(-\frac{\epsilon^2}{2\sigma_\theta^2L^2}\right)
\end{align*}

\end{document}